\documentclass{entcs}
\usepackage{entcsmacro}
\usepackage{graphicx}
\usepackage{amsmath, amsfonts}
\usepackage{epsfig}
\usepackage{url}
\usepackage{paralist}

\newcommand{\eqv}{\boldsymbol{=}\!\boldsymbol{=} \;}

\newcommand{\lar}{\boldsymbol{\Longleftarrow} \;}

\newcommand{\rar}{\boldsymbol{\Longrightarrow} \;}

\newcommand{\leqv}{\boldsymbol{\leq}\;\;}

\newcommand{\seq}{\; ; \;}

\newcommand{\br}{\displaybreak[0] \\}

\newcommand{\tensor}{\otimes}

\newcommand{\ket}[1]{|#1\rangle}
\newcommand{\bra}[1]{\langle#1|}
\newcommand{\braket}[2]{\langle #1 | #2 \rangle}

\newcommand{\ketb}[1]{|\textbf{#1}\rangle}

\def\lastname{Tafliovich}
\begin{document}
\begin{frontmatter}
  \title{Programming with Quantum Communication} 
  \author{Anya Tafliovich\thanksref{ALL}\thanksref{myemail}}
  \address{Computer Science\\ University of Toronto\\
    Toronto, Canada}
  \author{Eric C. R. Hehner\thanksref{coemail}}
  \address{Computer Science\\University of Toronto\\
    Toronto, Canada}
  \thanks[ALL]{This work is in part supported by NSERC.}
  \thanks[myemail]{Email:
    \href{mailto:anya@cs.toronto.edu}
    {\texttt{\normalshape anya@cs.toronto.edu}}}
  \thanks[coemail]{Email:
    \href{mailto:hehner@cs.toronto.edu}
    {\texttt{\normalshape hehner@cs.toronto.edu}}}

\begin{abstract}
  This work develops a formal framework for specifying, implementing,
  and analysing quantum communication protocols. We provide tools for
  developing simple proofs and analysing programs which involve
  communication, both via quantum channels and exhibiting the LOCC
  (local operations, classical communication) paradigm.
\end{abstract}
\begin{keyword}
  Quantum Computing, Quantum Communication Protocol, Formal
  Verification, Formal Methods of Program Design
\end{keyword}
\end{frontmatter}


\section{Introduction}
\label{sec:introduction}

The term quantum communication refers to the process of transferring a
quantum state between distinct physical locations. There are two ways
of accomplishing this task. The first one is analogous to classical
communication and involves sending a quantum bit over a quantum
communication channel (just as classical communication is associated
with sending classical bits over a classical communication channel).
The second one has no classical analogue. In a quantum world it is
possible to transfer a quantum bit without utilising a quantum
channel, by using a classical communication channel and a pair of
entangled states and applying quantum operations locally.

This work develops a formal framework for specifying, implementing,
and analysing quantum communication protocols. We provide tools for
developing simple proofs and analysing programs which involve
communication, both via quantum channels and exhibiting the LOCC
(local operations, classical communication) paradigm. We look at
quantum communication in the context of formal methods of program
development, or programming methodology. This is the field of computer
science concerned with applications of mathematics and logic to
software engineering tasks. In particular, the formal methods provide
tools to formally express specifications, prove correctness of
implementations, and reason about various properties of specifications
(e.g. implementability) and of implementations (e.g. time and space
complexity).

In this work the analysis of quantum communication protocols is based
on quantum predicative programming
(\cite{tafliovich06,tafliovich07,tafliovich04}), a recent
generalisation of the well-established predicative programming
(\cite{hehner:book,hehner04}).  It supports the style of program
development in which each programming step is proved correct as it is
made.  We inherit the advantages of the theory, such as its
generality, simple treatment of recursive programs, and of time and
space complexity. The theory of quantum programming provides tools to
write both classical and quantum specifications, develop quantum
programs that implement these specifications, and reason about their
comparative time, space, and communication complexity, all in the same
framework.

There has been a number of proposals for formal approaches to quantum
programming, including the language qGCL~\cite{sanders00,zuliani04},
process algebraic approaches of~\cite{adao,lalire04,jorrand04}, tools
developed in the field of category
theory~\cite{abramsky04,abramsky04:csqp,abramsky06,coecke04:le,selinger04},
functional languages
of~\cite{arrighi04,arrighi05,altenkirch05:fqpl,valiron04}, as well as
work of~\cite{dHondt06:wp},~\cite{danos07}, and~\cite{gay05}.  A
detailed discussion of the work related to quantum predicative
programming is presented in~\cite{tafliovich06}.

The contribution of this work is twofold. Firstly, we present a
framework for implementing quantum communication protocols, specifying
desired properties of the protocols, and formally proving whether
these properties hold. The properties are not restricted to reasoning
about the data sent or received by the parties involved. We provide
tools to prove properties which deal with the complexity of the
protocol, such as the number of classical and quantum bits sent during
its execution. Secondly, the reasoning about quantum communication
fits nicely in the general framework of quantum predicative
programming, and thus inherits all of its advantages. The definitions
of specification and program are simple: a specification is a boolean
(or probabilistic) expression and a program is a specification. The
treatment of recursion is simple: there is no need for additional
semantics of loops. The treatment of termination simply follows from
the introduction of a time variable; if the final value of the time
variable is $\infty$, then the program is a non-terminating one.
There is a uniform method for proving correctness and time, space, and
communication complexity; moreover, after proving them separately, we
naturally obtain the conjunction. The use of Dirac-like notation makes
it easy to write down specifications and develop algorithms. Finally,
the treatment of computation with mixed states does not require any
additional mechanisms.

The rest of this work is organised as follows.  Section~\ref{sec:qpp}
is a brief introduction to quantum predicative programming.  The
contribution of this work is Section~\ref{sec:distributedqs} which
introduces a formal framework for specifying, implementing, and
analysing quantum communication protocols and presents the analysis of
two such protocols: quantum teleportation and quantum dense coding.
Section~\ref{sec:conclusion} states conclusions and outlines
directions for future research. A short introduction to quantum
computing is presented in the Appendix \ref{sec:qc}.



\section {Quantum Predicative Programming}
\label{sec:qpp}

This section introduces the programming theory of our choice ---
quantum predicative programming. We briefly introduce parts of the
theory necessary for understanding Section~\ref{sec:distributedqs} of
this work. For a course in predicative programming the reader is
referred to~\cite{hehner:book}. An introduction to probabilistic
predicative programming can be found in~\cite{hehner04}. Quantum
predicative programming is developed
in~\cite{tafliovich06,tafliovich07,tafliovich04}.

\subsection{Predicative programming}
\label{sec:pp}

In predicative programming a specification is a boolean expression.
The variables in a specification represent the quantities of interest,
such as prestate (inputs), poststate (outputs), and computation time
and space. We use primed variables to describe outputs and unprimed
variables to describe inputs.  For example, specification $x' = x + 1$
states that the final value of $x$ is its initial value plus $1$.  A
computation \emph{satisfies} a specification if, given a prestate, it
produces a poststate, such that the pair makes the specification true.
A specification is \emph{implementable} if for each input state there
is at least one output state that satisfies the specification.

We use standard logical notation for writing specifications: $\wedge$
(conjunction), $\vee$ (disjunction), $\Rightarrow$ (logical
implication), $=$ (equality, boolean equivalence), $\neq$
(non-equality, non-equivalence), and \textbf{if then else}. The larger
operators $\eqv$, $\leqv$, and $\rar$ are the same as $=$, $\leq$, and
$\Rightarrow$, but with lower precedence. We use standard mathematical
notation, such as $ + \, - \, \times \, / \, mod \, div$.  We use
lowercase letters for variables of interest and uppercase letters for
specifications.

In addition to the above, we use the following notations: $\sigma$
(prestate), $\sigma'$ (poststate), $ok$ ($\sigma'=\sigma$), and $x:=e$
($x'=e \wedge y'=y \wedge \hdots$). The notation $ok$ specifies that
the values of all variables are unchanged. In the assignment $x:=e$,
$x$ is a state variable (unprimed) and $e$ is an expression (in
unprimed variables) in the domain of $x$.

If $R$ and $S$ are specifications in variables $x, y, \hdots \;$, then
the \emph{sequential composition} of $R$ and $S$ is defined by 
\begin{equation}
  \label{eq:seqcomp1}
  R\seq S \eqv \exists x'', y'', \hdots \cdot R'' \wedge S''
\end{equation}
where $R''$ is obtained from $R$ by substituting all occurrences of
primed variables $x', y', \hdots$ with double-primed variables $x'',
y'', \hdots \;$, and $S''$ is obtained from $S$ by substituting all
occurrences of unprimed variables $x, y, \hdots$ with double-primed
variables $x'', y'', \hdots \;$.

Various laws can be proved about sequential composition. One of the
most important ones is the substitution law, which states that for any
expression $e$ of the prestate, state variable $x$, and specification
$P$,
\begin{equation}
  \label{eq:substitution}
  x:=e\seq  P \eqv (\text{for } x \text{ substitute } e \text{ in } P)
\end{equation}

Specification $S$ \emph{is refined by} specification $P$ if and only
if $S$ is satisfied whenever $P$ is satisfied, that is $\forall
\sigma, \sigma' \cdot S \Leftarrow P$. Given a specification, we are
allowed to implement an equivalent specification or a stronger one.

A \emph{program} is an implemented specification. A good basis for
classical (non-quantum) programming is provided by: $ok$, assignment,
\textbf{if then else}, sequential composition, booleans, numbers,
bunches, and functions. Given a specification $S$, we proceed as
follows. If $S$ is a program, there is no work to be done. If it is
not, we build a program $P$, such that $P$ refines $S$, i.e. $S
\Leftarrow P$. The refinement can proceed in steps: $S \Leftarrow
\hdots \Leftarrow R \Leftarrow Q \Leftarrow P$.

In $S \Leftarrow P$ it is possible for $S$ to appear in $P$. No
additional rules are required to prove the refinement. For example, it
is trivial to prove that
\begin{equation*}
  x \geq 0 \Rightarrow x'=0 
  \lar
  \textbf{if } x=0
  \textbf{ then } ok 
  \textbf{ else } (x:=x-1\seq  x \geq 0 \Rightarrow x'=0 )
\end{equation*}

The specification says that if the initial value of $x$ is
non-negative, its final value must be $0$. The solution is: if the
value of $x$ is zero, do nothing, otherwise decrement $x$ and repeat.

\subsection{Probabilistic predicative programming}
\label{sec:ppp}

A \emph{probability} is a real number between $0$ and $1$, inclusive.
A \emph{distribution} is an expression whose value is a probability
and whose sum over all values of variables is $1$. Given a
distribution of several variables, we can sum out some of the
variables to obtain a distribution of the rest of the variables.

To generalise boolean specifications to probabilistic specifications,
we use $1$ and $0$ both as numbers and as boolean $\mathit{true}$ and
$\mathit{false}$, respectively. If $R$ and $S$ are specifications in
variables $x, y, \hdots \;$, then the definition (\ref{eq:seqcomp1})
of \emph{sequential composition} of $R$ and $S$ is generalised to
\begin{equation*}
  R\seq S \eqv \sum {x'', y'', \hdots} \cdot R'' \times S''
\end{equation*}
where $R''$ and $S''$ are defined as before.

If $p$ is a probability and $R$ and $S$ are distributions, then
\begin{equation*}
  \textbf{if } p \textbf{ then } R \textbf{ else } S \eqv
  p \times R + (1-p)\times S
\end{equation*}

If $S$ is an implementable deterministic specification and $p$ is a
distribution of the initial state $x, y, ...$, then the distribution
of the final state is
\begin{equation*}
  p'\seq  S
\end{equation*}

Various laws can be proved about sequential composition. One of the
most important ones, the substitution law, introduced earlier, applies
to probabilistic specifications as well.

\subsection{Quantum Predicative Programming}
\label{sec:qpp_sub}

Let $\mathbb{C}$ be the set of all complex numbers with the absolute
value operator $|\cdot|$ and the complex conjugate operator $^*$. Then
a state of an $n$-qubit system is a function $\psi : 0,..2^n
\rightarrow \mathbb{C}$, such that $\sum {x:0,..2^n} \cdot |\psi x|^2
\eqv 1$.  Here notation $i,..j$ means from (and including) $i$ to (and
excluding) $j$.

If $\psi$ and $\phi$ are two states of an $n$-qubit system, then their
\emph{inner product}, denoted by $\braket{\psi}{\phi}$, is
defined by:
\begin{equation*}
  \braket{\psi}{\phi} = \sum {x:0,..2^n} \cdot (\psi x)^* \times (\phi x)
\end{equation*}

A \emph{basis} of an $n$-qubit system is a collection of $2^n$ quantum
states $b_{0,..2^n}$, such that $\forall i,j:0,..2^n \cdot
\braket{b_i}{b_j} = (i=j)$. We adopt the following Dirac-like notation
for the computational basis: if $x$ is from the domain $0,..2^n$, then
$\textbf{x}$ denotes the corresponding $n$-bit binary encoding of $x$
and $\ketb{x}: 0,..2^n \rightarrow \mathbb{C}$ is the following
quantum state:
\begin{equation*}
  \ketb{x} = \lambda i:0,..2^n \cdot (i=x)
\end{equation*}
where $\lambda x: D \cdot b$ is a function of a variable $x$ with
domain $D$ and body $b$.  If $\psi$ is a state of an $m$-qubit system
and $\phi$ is a state of an $n$-qubit system, then $\psi \tensor
\phi$, the tensor product of $\psi$ and $\phi$, is the following state
of a composite $m+n$-qubit system:
\begin{equation*}
  \psi \tensor \phi = \lambda i:0,..2^{m+n} \cdot 
                      \psi (i\:div\:2^n) \times \phi (i\:mod\:2^n)
\end{equation*}

We write $\phi^{\tensor n}$ to mean ``$\phi$ tensored with itself $n$
times''.  An operation defined on an $n$-qubit quantum system is a
higher-order function, whose domain and range are maps from $0,..2^n$
to the complex numbers. An \emph{identity} operation on a state of an
$n$-qubit system is defined by
\begin{equation*}
  I^n = \lambda \psi : 0,..2^n \rightarrow \mathbb{C} \cdot \psi
\end{equation*}

For a linear operation $A$, the \emph{adjoint} of $A$, written
$A^\dagger$, is the (unique) operation, such that for any two states
$\psi$ and $\phi$, $\braket{\psi}{A \phi} = \braket{A^{\dagger}
  \psi}{\phi}$.

The \emph{unitary transformations} that describe the evolution of an
$n$-qubit quantum system are operations $U$ defined on the system,
such that $U^{\dagger} U = I^n$.

In this setting, the \emph{tensor product} of operators is defined in
the usual way. If $\psi$ is a state of an $m$-qubit system, $\phi$ is
a state of an $n$-qubit system, and $U$ and $V$ are operations defined
on $m$ and $n$-qubit systems, respectively, then the tensor product of
$U$ and $V$ is defined on an $m+n$ qubit system by 
$$(U \tensor V) (\psi \tensor \phi) = (U \psi) \tensor (V \phi)$$

To apply an operation $U$ defined on a $1$-qubit system to qubit $i$
in a composite $n$-qubit system, we apply the operation $U_i^{n}$ to
the entire system, where $U_i^{n}$ is defined by:
\begin{equation*}
  U_i^n = \underset{i}{\underbrace{I \tensor ... \tensor I}}
          \:\tensor\: U \tensor\:
          \underset{n-i-1}{\underbrace{I \tensor ... \tensor I}}
\end{equation*}

Suppose we have a system of $n$ qubits in state $\psi$ and we measure
(observe) it. Suppose also that we have a variable $r$ from the domain
$0,..2^n$, which we use to record the result of the measurement, and
variables $x,y, \hdots$, which are not affected by the measurement.
Then the measurement corresponds to a probabilistic specification that
gives the probability distribution of $\psi'$ and $r'$ (these depend
on $\psi$ and on the type of measurement) and states that the
variables $x,y,\hdots$ are unchanged.

For a general quantum measurement described by a collection $M =
M_{0,..2^n}$ of measurement operators, which satisfy the completeness
equation (see Appendix \ref{sec:qc}), the specification is
$\textbf{measure}_M \, \psi \, r$, where
\begin{equation*}
  \textbf{measure}_M \, \psi \, r \eqv
    \braket{\psi}{M_{r'}^\dagger M_{r'}\psi} \times
    \left(
     \psi'=\frac {M_{r'}\psi}{\sqrt{\braket{\psi}
                 {M_{r'}^\dagger M_{r'}\psi}}}
    \right) \times (\sigma'=\sigma)
\end{equation*}
where $\sigma'=\sigma$ is an abbreviation of $(x'=x) \times (y'=y)
\times \hdots$ and means ``all other variables are unchanged''.

Given an arbitrary orthonormal basis $B = b_{0,..2^n}$, measurement of
$\psi$ in basis $B$ is:
\begin{equation*}
  \textbf{measure}_B \, \psi \, r \eqv 
    |\braket{b_{r'}}{\psi}|^2 \times (\psi'=b_{r'}) \times
      (\sigma'=\sigma)
\end{equation*}

The simplest and the most commonly used measurement in the
computational basis is:
\begin{equation*}
  \textbf{measure } \psi \, r \eqv
    | \psi r' |^2  \times (\psi'= |\textbf{r}'\rangle) \times (\sigma'=\sigma)
\end{equation*}

In this case the distribution of $r'$ is $|\psi r'|^2$ and the
distribution of the quantum state is:
\begin{equation*}
  \sum {r'} \cdot |\psi r'|^2 \times (\psi' = |\textbf{r}'\rangle)
\end{equation*}
which is precisely the mixed quantum state that results from the
measurement. 

In order to develop quantum programs we need to add to our list of
implemented things. We add variables of type quantum state as above
and we allow the following three kinds of operations on these
variables.  If $\psi$ is a state of an $n$-qubit quantum system, $r$
is a natural variable, and $M$ is a collection of measurement
operators that satisfy the completeness equation, then:
\begin{enumerate}
\item $\psi:=\ket{0}^{\tensor n}$ is a program
\item $\psi:=U\psi$, where $U$ is a unitary transformation on an
  $n$-qubit system, is a program
\item $\textbf{measure}_M \, \psi \, r$ is a program
\end{enumerate}
where the superscript $^{\tensor{n}}$ means ``tensored with itself $n$
times''. The special cases of measurements are therefore also allowed.

Some unitary operations that we will use in the later sections are
(here $x,c:0,1$):
\begin{align*}
  & I \ket{x} = \ket{x} &\text{identity}\br
  & X \ket{x} = \ket{1-x} &\text{X - Pauli matrix}\br
  & Y \ket{x} = (-1)^{x} \times i \times \ket{1-x} &\text{Y - Pauli matrix}\br
  & Z \ket{x} = (-1)^{x} \times \ket{x} &\text{Z - Pauli matrix}\br
  & H \ket{x} = (\ket{0} + (-1)^{x} \times \ket{1}) / \sqrt{2} 
   &\text{Hadamard}\br
  & CNOT \ket{cx} = (I \tensor X^{c}) \ket{cx} &\text{controlled-not}
\end{align*}



\section{Distributed Quantum Systems and Communication}
\label{sec:distributedqs}

In predicative programming, to reason about distributed computation we
(disjointly) partition the variables between the processes involved in
a computation. Parallel composition is then simply boolean conjunction.
For example, consider two processes $P$ and $Q$. $P$ owns integer
variables $x$ and $y$ and $Q$ owns an integer variable $z$. Suppose
$ P \eqv x:=x+1\seq  y:=x \text{ and } Q \eqv z:=-z $.
Parallel composition of $P$ with $Q$ is then simply
\begin{align*}
  & P || Q \eqv P \land Q 
  \eqv
   x'=x+1 \land y'=x+1 \land z'=-z
\end{align*}

In quantum predicative programming, one needs to reason about
distributed quantum systems. Recall that if $\psi$ is a state of an
$m$-qubit system and $\phi$ is a state of an $n$-qubit system, then
$\psi \tensor \phi$, the tensor product of $\psi$ and $\phi$, is the
state of a composite $m+n$-qubit system. On the other hand, given a
composite $m+n$-qubit system, it is not always possible to describe it
in terms of the tensor product of the component $m$- and $n$-qubit
systems. Such a composed system is \emph{entangled}. Entanglement is
one of the most non-classical, most poorly understood, and most
interesting quantum phenomena. An entangled system is in some sense
both distributed and shared. It is distributed in the sense that each
party can apply operations and measurements to only its qubits. It is
shared in the sense that the actions of one party affect the outcome of
the actions of another party. Simple partitioning of qubits is
therefore insufficient to reason about distributed quantum
computation. 

The formalism we introduce fully reflects the physical properties of a
distributed quantum system. We start by partitioning the qubits
between the parties involved. For example, consider two parties $P$
and $Q$. $P$ owns the first qubit of the composite entangled quantum
system $\psi = \ket{00}/\sqrt{2} + \ket{11}/\sqrt{2}$ and $Q$ owns the
second qubit. A specification is a program only if each party computes
with its own qubits. In our example, 
$$P \eqv \psi_0 := H \psi_0\seq \textbf{measure } \psi_0\; p
\quad\text{and}\quad Q \eqv \textbf{measure } \psi_1\; q$$ are
programs, if $p$ and $q$ are integer variables owned by $P$ and $Q$,
respectively. Note that we cannot write down expressions for $\psi_0$
and $\psi_1$: this is consistent with the laws of quantum mechanics as
$\psi$ is an entangled state. Parties $P$ and $Q$ can access only
their own qubits: they could in theory be light years apart.

Sometimes we want to explicitly include partitioning of variables as
part of a specification. For this purpose, we introduce notation
$\textbf{var}_P$ to mean the bunch of variables that belong to process
$P$. In the above example we can make the partitioning of variables
explicit with the specification
$$ \psi_0, p : \textbf{var}_P \land \psi_1, q : \textbf{var}_Q $$

We define parallel composition of $P$ and $Q$ which share an $n+m$
quantum system in state $\psi$ with the first $n$ qubits belonging to
$P$ and the other $m$ qubits belonging to $Q$ as follows. If
$$ P \eqv \psi_{0,..n} := U_P \psi_{0,..n} 
   \quad\text{and}\quad
   Q \eqv \psi_{n,..n+m} := U_Q \psi_{n,..n+m}$$
where $U_P$ is a unitary operation on an $n$-qubit system and $U_Q$ is
a unitary operation on an $m$-qubit system, then
$$ P\; ||_\psi\; Q \eqv \psi := (U_P \tensor U_Q) \psi$$

Performing $ok$ is equivalent to performing the identity unitary
operation, and therefore if
$$ P \eqv \psi_{0,..n} := U_P \psi_{0,..n} 
   \quad\text{and}\quad
   Q \eqv ok$$
then
$$ P\; ||_\psi\; Q \eqv \psi := (U_P \tensor I^{\tensor m}) \psi$$

Similarly, if
$$ P \eqv \textbf{measure}_{M_P}\; \psi_{0,..n} \;p 
   \quad\text{and}\quad
   Q \eqv \textbf{measure}_{M_Q}\; \psi_{n,..n+m} \;q$$
where $M_P$ and $M_Q$ are a collection of proper measurement operators
for $n$- and $m$-qubit systems, respectively, then
$$P\; ||_\psi\; Q \eqv \textbf{measure}_{M_P \tensor M_Q} \psi \; pq$$
where $pq$ is the number that corresponds to the binary string
\textbf{pq}. 

In our example,
\begin{alignat*}{2}
  & \psi := \ket{00}/\sqrt{2} + \ket{11}/\sqrt{2}\seq
     P \;||_\psi Q 
   &\text{expand, substitute}\br
\eqv
  & \psi := \ket{00}/\sqrt{2} + \ket{11}/\sqrt{2}\seq  &\\
   & \textbf{measure } (H \psi_0)\; p 
     \;||_\psi\;
    \textbf{measure } \psi_1\; q
   &\text{compose on }\psi\br
\eqv
  & \psi := \ket{00}/\sqrt{2} + \ket{11}/\sqrt{2}\seq
    \textbf{measure } (H \tensor I) \psi \; pq
   &\text{substitute}\br
\eqv
  & \textbf{measure } (H \tensor I) 
      (\ket{00}/\sqrt{2} + \ket{11}/\sqrt{2}) \; pq
   &\text{apply }H \tensor I\br
\eqv
  & \textbf{measure }  
      (\ket{00} + \ket{01} + \ket{10} - \ket{11})/2 \; pq
   &\text{measure}\br
\eqv
  & \left|(\ket{00} + \ket{01} + \ket{10} - \ket{11})/2 \; pq \right|^2
    \times
    (\psi' = |\textbf{p}'\textbf{q}'\rangle)
   &\text{application}\br
\eqv
  & (\psi'= |\textbf{p}'\textbf{q}'\rangle)/4
\end{alignat*}

When explicitly specifying partitioning of variables in a parallel
composition, it is convenient to allow the variables to appear as
subscripts on the corresponding processes. For example, the
specification
$ P_{\psi_0,p} \;||_\psi\; Q_{\psi_1,q} $
denotes a parallel composition of processes $P$ and $Q$ that share an
entangled state $\psi$, such that $\psi_0$ and $p$ belong to $P$ and
$\psi_1$ and $q$ belong to $Q$.

To reason about communication between processes we use the framework
of Hehner's calculus(\cite{hehner:book}). A named, one-way
communication channel $c$ is described by an infinite message script
$M_c$, an infinite time script $T_c$, and read and write cursors $r_c$
and $w_c$. The message and time scripts are the list of all messages
that appear on the channel and the list of corresponding times. The
read and write cursors specify how many messages have been read from
and written to a channel. To specify two-way communication, we use two
channels. The input and output on channel $c$ are defined by the
following operations (here $t$ is the time variable):
\begin{align*}
  & c! e &\eqv &M_c w_c = e \land T_c w_c = t \land w_c:=w_c+1
      &\text{$c$ output $e$}\\
  & c?   &\eqv &r_c:=r_c+1  &\text{$c$ input}\\
  & c    &\eqv &M_c (r_c-1)  &
\end{align*}

A channel declaration $ \textbf{chan } c:T \cdot P$ defines a new
channel $c$ with communication of type $T$; the declaration applies to
the specification $P$ (here $xnat$ stands for naturals extended with
$\infty$):
$$ \textbf{chan } c:T \cdot P \eqv \exists M_c:[\infty*T] \cdot
\exists T_c:[\infty*real] \cdot \textbf{var } r_c, w_c:xnat := 0 \cdot
P$$ where $[\infty*T]$ is an infinite sequence of elements of type
$T$. One useful theorem that we use in later examples is the
equivalence of communication on a local channel with assignment:
$$ \textbf{chan } c:T \cdot c! e \;||\; (c? \seq x:=c) \eqv x:=e $$ 
The reader is referred to \cite{hehner:book} for a detailed
description of formal treatment of classical communication in Hehner's
calculus.

When defining a quantum communication channel one must be careful not
to introduce any unwanted behaviour, such as violation of the
no-cloning principle (i.e. creation of identical copies of an
unknown arbitrary quantum state). For this purpose we make the change
of ownership of the transported qubit explicit in the definition:
\begin{align*}
  & c! \psi &\eqv &M_c w_c = \psi \land T_c w_c = t \land w_c'=w_c+1 
    \land
   \textbf{var}_P' = \textbf{var}_P \backslash \psi \land \sigma' = \sigma\\
  & c? \psi &\eqv & r_c'=r_c+1 \land \psi' = M_c r_c \land 
             \textbf{var}_Q' = \textbf{var}_Q,\psi \land \sigma' = \sigma
\end{align*}
where $c$ is a quantum communication channel from process $P$ to
process $Q$ and $\sigma' = \sigma$ is shorthand for ``the rest of the
variables are unchanged''.

Now that we allow changing of ownership of the variables, the
specification $\sigma_P'=\sigma_P$, ``the rest of the variables of
process $P$ are unchanged'' is defined by $\forall v':\textbf{var}_P'
\cdot v'=v$.

The declaration of a quantum channel $\textbf{qchan } c: qbit \cdot P$
is similar to the declaration of a local classical channel:
$$ \textbf{qchan } q:T \cdot P \eqv \exists M_q:[\infty*T] \cdot
\exists T_q:[\infty*real] \cdot \textbf{var } r_q, w_q:xnat := 0 \cdot
P$$

Similarly to the above-mentioned theorem, we can prove the equivalence
of communication on a local quantum channel with the change of
ownership.  If $P \eqv c!\psi$ and $Q \eqv c? \psi$, then (leaving out
time)
\begin{align*}
  & \textbf{qchan } c:qbit \cdot P \; || \; Q 
   &\text{def. qchan}\br
\eqv
  & \exists M : [\infty * qbit] \cdot \textbf{var } r,w : xnat := 0 \cdot 
    P \; || \; Q 
   &\text{expand}\br
\eqv
  & \exists M : [\infty * qbit] \cdot \textbf{var } r,w : xnat := 0\cdot\\ 
    &\quad  M w = \psi \land w'=w+1 \land 
            \textbf{var}_P' = \textbf{var}_P \backslash \psi\\
    &\quad \land \textbf{var}_Q' = \textbf{var}_Q,\psi \land 
      r'=r+1 \land \psi' = Mr
  &\text{initialisation} \br
\eqv
  & \exists M : [\infty * qbit] \cdot \textbf{var } r,w : xnat\cdot\\ 
    &\quad  M 0 = \psi \land w'=1 \land 
            \textbf{var}_P' = \textbf{var}_P \backslash \psi\\
    &\quad \land \textbf{var}_Q' = \textbf{var}_Q,\psi \land 
      r'=1 \land \psi' = M0
  &\text{simplify} \br
\eqv
  & \textbf{var}_P' = \textbf{var}_P \backslash \psi \land
    \textbf{var}_Q' = \textbf{var}_Q,\psi \land
    \sigma' = \sigma
\end{align*}


\subsection{Quantum teleportation}
\label{sec:teleportation}

Quantum teleportation is the most famous quantum communication
protocol. Its description first appeared in a seminal article by
Bennett {\it et al} in 1993 (\cite{bennett93}), it has since been
extensively used as part of more complex quantum communication
protocols, and has received much attention in experimental research.
The protocol achieves transmission of quantum information by utilising
only a classical communication channel and an entangled pair of
qubits: no qubits are sent in the process.

\textbf{The protocol}: Alice and Bob share an entangled pair of qubits
in the state $(\ket{00}+\ket{11})/\sqrt{2}$. Alice has some qubit
$\psi$ in her possession (she may not know the state of the qubit) that
she wishes to transfer to Bob. Alice starts by interacting the qubit
she wishes to teleport with her half of the entangled pair (she
applies a controlled-not followed by a Hadamard transform) and
measuring her two qubits. She then sends the results of her
measurements to Bob (two classical bits). Bob receives the two
classical bits and, depending of their values, applies one of the
three Pauli operators or the identity to his qubit. Surprisingly, he
has recovered the state Alice wished to teleport.

\begin{figure}[t]
  \centering
  \epsfig{figure=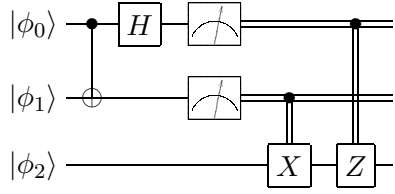}
  \caption{Quantum teleportation protocol}
  \label{fig:teleportation}
\end{figure}

The protocol is usually described informally, by using a diagram as in
Figure~\ref{fig:teleportation}\footnote{The figure is generated with
  {\it qasm2pdf}}. Such a description is insufficient, in part since
it only describes the evolution of the quantum system and does not
specify the distribution of the system nor the communication.
Alternatively, the description of the protocol is given informally, in
English. Our goal is to formally define and prove correctness of the
quantum teleportation protocol. Some approaches proposed in the
literature (e.g.~\cite{zuliani:thesis}) define teleportation as a
program that implements a specification of the form $\phi' = \psi$. We
point out that this specification may as well be implemented by a
program that involves sending a qubit on a quantum channel, which is
not teleportation.  Furthermore, the specification does not mention
that two classical bits are sent on a classical channel, which is an
important part of the specification of teleportation.  Similarly, it
is important to specify that a pair of maximally entangled qubits is
required.

To formalise the quantum teleportation protocol we let $c$ be the
number of classical bits sent on a communication channel and $q$ be
the number of quantum bits sent. The formal specification of quantum
teleportation is:
\begin{align*}
  S \eqv\quad
  &\phi_{01} : \textbf{var}_{Alice} \land \phi_2 : \textbf{var}_{Bob} \; \land\\ 
  &\phi_{0,..3} = (\alpha \times \ket{0} + \beta \times \ket{1})
                  \tensor
                  (\ket{00} + \ket{11})/\sqrt{2} \\
  \Rightarrow\;
  &\phi_2' = \alpha \times \ket{0} + \beta \times \ket{1} \land c'=c+2 \land q'=q
\end{align*}

The specification says that if the computation starts with a qubit
(specified in the most general form) in Alice's possession and if
Alice and Bob share a maximally entangled state
$(\ket{00}+\ket{11})/\sqrt{2}$, then at the end of the computation the
qubit is teleported to Bob at a cost of 2 classical bits of
communication and 0 qubits of communication. The specification does
not restrict the quantum system to three qubits, so that teleportation
can be a part of a bigger computation.

The well-known solution is the following parallel program:
\begin{align*}
  P \eqv 
  & \textbf{chan } ch: bit \cdot 
      Alice_{a_0,a_1,\phi_{01}}
      \; ||_{\phi} \;
      Bob_{b_0,b_1,\phi_2} \\
    \text{where }
  & Alice \eqv 
  \begin{aligned}[t]
   & \phi_{01} := CNOT \phi_{01}\seq \phi_0 := H \phi_0\seq 
     \textbf{measure } \phi_{01} \; a_0a_1\seq \\
   & c:=c+1\seq ch!a_0\seq c:=c+1\seq ch!a_1
 \end{aligned}\\
 \text{and }
  & Bob \eqv ch? \seq  b_0 := ch \seq  ch? \seq  b_1 := ch \seq 
             \phi_2 := Z^{b_0} X^{b_1} \phi_2
\end{align*}

That is, two processes, $Alice$ and $Bob$, partition a 3-qubit quantum
system $\phi$, so that $Alice$ owns the first qubit (the one she wants
to teleport) and the second qubit and $Bob$ owns the third qubit.
$Alice$ can write to a local classical communication channel $ch$ and
$Bob$ can read from it. Finally, $a_0$ and $a_1$ are two bits that
belong to $Alice$, and $b_0$ and $b_1$ are two bits that belong to
$Bob$. The process $Alice$ performs local operations and measurements
and sends two classical bits on the channel. The process $Bob$ reads
from the channel and performs local operations.

Our goal is to prove that the program $P$ implements the specification
$S$. We first note the following equivalence:
\begin{align*}
  & P \Rightarrow S 
   &\text{def. } S\br
\eqv\quad 
  &P \land 
      \phi_{01} : \textbf{var}_{Alice} \land 
       \phi_2 : \textbf{var}_{Bob} \; \land \\
     & \phi_{0,..3} = (\alpha \times \ket{0} + \beta \times \ket{1})
                       \tensor
                       (\ket{00} + \ket{11})/\sqrt{2}\\
  \Rightarrow\;
  & \phi_2' = \alpha \times \ket{0} + \beta \times \ket{1}
    \land c'=c+2 \land q'=q
  &\text{simplification}\br
\eqv\quad
  & P \land \phi = (\alpha \times \ket{0} + \beta \times \ket{1})
                   \tensor
                   (\ket{00} + \ket{11})/\sqrt{2} \\
  \Rightarrow\;
  & \phi_2' = \alpha \times \ket{0} + \beta \times \ket{1}
    \land c'=c+2 \land q'=q
\end{align*}

Next, we simplify $P$ to prove the above implication. With implicit
partitioning of variables (as it does not change):
\mathindent=10pt
\begin{align*}
  & \textbf{chan } ch: bit \cdot
  \begin{aligned}[t]
    ((&\phi_{01} := CNOT \phi_{01}\seq \phi_0 := H \phi_0\seq 
       \textbf{measure } \phi_{01} \; a_0a_1\seq \\
      &c:=c+1\seq ch!a_0\seq c:=c+1\seq ch!a_1) \\
    ||_{\phi}\; 
     (&ch? \seq  b_0 := ch \seq  ch? \seq  b_1 := ch \seq 
       \phi_2 := Z^{b_0} X^{b_1} \phi_2))
  \end{aligned}\br
&\hspace{-3mm}\eqv\hspace{5cm}\text{substitute, H on first qubit}\br
  & \textbf{chan } ch: bit \cdot
  \begin{aligned}[t]
    ((&\textbf{measure } H\tensor I(CNOT \phi_{01}) \; a_0a_1 \seq\\ 
      &c:=c+1\seq ch!a_0\seq c:=c+1\seq ch!a_1) \\
    ||_{\phi}\;
     (&ch? \seq  b_0 := ch \seq  ch? \seq  b_1 := ch \seq 
       \phi_2 := Z^{b_0} X^{b_1} \phi_2))
  \end{aligned}\br
&\hspace{-3mm}\eqv\hspace{5cm}\text{parallel composition, simplification}\br
  &\textbf{chan } ch: bit \cdot\\
  &\;\;\textbf{measure}_{01}\; H\tensor I\tensor I (CNOT\tensor I \phi) \; a_0a_1\seq \\
  &\;\;((c:=c+1\seq  ch!a_0\seq c:=c+1\seq ch!a_1) 
          \;||\; (ch? \seq  b_0 := ch \seq  ch? \seq  b_1 := ch)) \seq \\
  &\;\;\phi := I \tensor I \tensor Z^{b_0} (I \tensor I \tensor X^{b_1}\phi)\br
&\hspace{-3mm}\eqv\hspace{6cm}\text{classical channel}\br
   &\;\;\textbf{measure}_{01}\; H\tensor I\tensor I (CNOT\tensor I \phi) \; a_0a_1\seq\\
   &\;\;c'=c+2 \land b_0'=a_0 \land b_1'=a_1 \land \sigma'=\sigma\seq \\
   &\;\;\phi := I \tensor I \tensor Z^{b_0} (I \tensor I \tensor X^{b_1} \phi)
\end{align*}

Next we notice that the first line in the above specification (which
is, in fact, the effect of Alice's actions) conjoined with the
specification of the initial state of the quantum system, result in
the following distribution over the states of the computation:
\begin{align*}
  & \phi' = \ket{a_0'a_1'} \tensor 
  (\alpha \times \ket{a_1'} + (-1)^{a_0'} \times \beta \times \ket{1-a_1'})/4 
\end{align*}

That is, with probability $1/4$ the quantum system is in state
$\ket{00} \tensor (\alpha \times \ket{0} + \beta \times \ket{1})$ and
the values of Alice's bits are $a_0=0$ and $a_1=0$; with probability
$1/4$ the quantum system is in state $\ket{01} \tensor (\alpha \times
\ket{1} + \beta \times \ket{0})$ and the values of Alice's bits are
$a_0=0$ and $a_1=1$; etc. 

To prove this formally, we first note that:
%
\begin{align*}
 &&& H \tensor I \tensor I (CNOT \tensor I 
         ((\alpha \times \ket{0} + \beta \times \ket{1})
           \tensor
           (\ket{00} + \ket{11})/\sqrt{2}))\br
&\eqv&  &\hspace{8cm}\text{apply CNOT}\\
 &&& H \tensor I \tensor I (\alpha \times \ket{000} + \beta \times \ket{110} + 
   \alpha \times \ket{011} + \beta \times \ket{101} ) / \sqrt{2}\br
&\eqv&&\hspace{8cm}\text{apply H}\\
 &&& \alpha \times (\ket{0} + \ket{1}) \tensor \ket{00} / 2 +
   \beta \times (\ket{0} - \ket{1}) \tensor \ket{10} / 2 +\\
 &&& \alpha \times (\ket{0} + \ket{1}) \tensor \ket{11} / 2 +
   \beta \times (\ket{0} - \ket{1}) \tensor \ket{01} / 2\br 
&\eqv&&\hspace{8cm}\text{rearrange terms}\\
 &&& \ket{00} \tensor (\alpha \times \ket{0} + \beta \times \ket{1}) /2 +
   \ket{01} \tensor (\alpha \times \ket{1} + \beta \times \ket{0}) /2 +\\
 &&& \ket{10} \tensor (\alpha \times \ket{0} - \beta \times \ket{1}) /2 +
   \ket{11} \tensor (\alpha \times \ket{1} - \beta \times \ket{0}) /2
\end{align*}

Therefore, measurement of the first two qubits of the above state in
the computational basis gives:
\begin{align*}
  & \;\;\textbf{measure}_{01}\; 
    \begin{aligned}[t]
      & (\ket{00} \tensor (\alpha \times \ket{0} + \beta \times \ket{1}) /2 +
         \ket{01} \tensor (\alpha \times \ket{1} + \beta \times \ket{0}) /2 +\\
      & \;\ket{10} \tensor (\alpha \times \ket{0} - \beta \times \ket{1}) /2 +
         \ket{11} \tensor (\alpha \times \ket{1} - \beta \times \ket{0}) /2)\\
      &a_0a_1\\
    \end{aligned}\\
&\eqv
   \phi' = \ket{a_0'a_1'} \tensor 
       (\alpha \times \ket{a_1'} + (-1)^{a_0'} \times \beta \times \ket{1-a_1'})/4 
\end{align*}

Let $Q$ be the specification of the initial state of the quantum
system:
\begin{equation*}
 Q \eqv \phi = (\alpha \times \ket{0} + \beta \times \ket{1})
                   \tensor
                   (\ket{00} + \ket{11})/\sqrt{2}
\end{equation*}

Putting it all together, we get:
\begin{align*}
  & Q \land P 
  &\text{}\br
\eqv
  & Q \times 
   (\textbf{measure}_{01}\; H\tensor I\tensor I (CNOT\tensor I \phi) \; a_0a_1\seq \\
   & c'=c+2 \land b_0'=a_0 \land b_1'=a_1 \land \sigma'=\sigma\seq 
   &\hspace{-3mm}\text{one point law,}\\
   & \phi := I \tensor I \tensor Z^{b_0} (I \tensor I \tensor X^{b_1} \phi))
   &\text{as above}\br
\eqv
  & Q \times  \phi' = \ket{a_0'a_1'} \tensor 
       (\alpha \times \ket{a_1'} + (-1)^{a_0'} \times \beta \times\ket{1-a_1'})/4\seq 
   &\text{sequential}\\
   & c'=c+2 \land b_0'=a_0 \land b_1'=a_1 \land \sigma'=\sigma\seq 
   &\text{composition,}\\
   & \phi := I \tensor I \tensor Z^{b_0} (I \tensor I \tensor X^{b_1} \phi)
   &\text{one point law}\br
\eqv
  & Q \times (c'=c+2) \times (b_0'=a_0') \times (b_1'=a_1') \times
  (\sigma'=\sigma) \times \\
   & \phi' = I \tensor I \tensor Z^{b_0'} (I \tensor I \tensor X^{b_1'} \\
   & \quad   (\ket{b_0'b_1'} \tensor 
              (\alpha \times \ket{b_1'} + 
               (-1)^{b_0'} \times \beta \times\ket{1-b_1'})/4))
   &\text{apply $X^{b_1'}$}\br
\eqv
  & Q \times (c'=c+2) \times (b_0'=a_0') \times (b_1'=a_1') \times 
            (\sigma'=\sigma) \times\\
   & \phi' = I \tensor I \tensor Z^{b_0'}
             (\ket{b_0'b_1'} \tensor 
              (\alpha \times \ket{0} + 
               (-1)^{b_0'} \times \beta \times\ket{1})/4
   &\text{apply $Z^{b_0'}$}\br
\eqv
  & Q \times (c'=c+2) \times (b_0'=a_0') \times (b_1'=a_1') \times 
         (\sigma'=\sigma) \times\\
   & (\phi' = \ket{b_0'b_1'} \tensor 
              (\alpha \times \ket{0} + \beta \times\ket{1})/4)
   &\text{}\br
\leqv
 & (c'=c+2) \times (q'=q) \times 
   (\phi_2' = \alpha \times \ket{0} + \beta \times\ket{1})\br
\eqv 
  &S
\end{align*}
where $\leqv$ is the generalisation of $\rar$ for probabilistic
specification (see~\cite{hehner04}). 

This example shows formalisation and analysis of an LOCC (local
operations, classical communication) quantum communication protocol.
We now turn to our attention to a protocol which involves a quantum
communication channel.


\subsection{Quantum dense coding}
\label{sec:densecoding}
 
The quantum dense coding (sometimes called super-dense coding) protocol
is less famous than the quantum teleportation protocol, but it is no
less important. It achieves the transfer of 2 bits of classical
information by sending 1 bit of quantum information and utilising 1
entangled pair of qubits. That is, its goal is the opposite of that
of the quantum teleportation protocol. 

Just as with teleportation, the protocol is usually described
informally: either with a diagram or in English. We formalise the
specification of the protocol by using the same variables as in
section~\ref{sec:teleportation}:
\begin{align*}
   S \eqv \quad& a_0, a_1, \phi_0 : \textbf{var}_{Alice} \land
            b_0, b_1, \phi_1 : \textbf{var}_{Bob} \land
            \phi_{01} = (\ket{00} + \ket{11})/\sqrt{2}\\
           \Rightarrow\;
          & b_0'=a_0 \land b_1'=a_1 \land c'=c \land q'=q+1
\end{align*}

The specification says that if the computation starts with Alice and
Bob sharing a maximally entangled state, with classical bits $a_0$ and
$a_1$ in Alice's possession and $b_0$ and $b_1$ in Bob's possession,
then at the end of the computation Bob has the values of Alice's
classical bits, at a cost of sending no bits on a classical channel
and one qubit on a quantum channel. The program for the protocol is:
\begin{align*}
  P \eqv & \textbf{qchan } qch : qbit \cdot 
             Alice_{a_0,a_1,\phi_0} \; ||_{\phi}\;  Bob_{b_0,b_1,\phi_1}\\
  \text{where } 
   Alice \eqv & \textbf{if } a_0=a_1=0 \textbf{ then } ok \\
             & \textbf{else if } a_0=0 \land a_1=1 \textbf{ then } \phi_0:=X\phi_0 \\
             & \textbf{else if } a_0=1 \land a_1=0 \textbf{ then } \phi_0:=Z\phi_0 \\
             & \textbf{else } \phi_0:=Y\phi_0\seq  \\
             & q:=q+1\seq  qch! \phi_0\\
  \text{and }
  Bob \eqv  & qch? \phi_0\seq  \phi:=CNOT\phi\seq  \phi_0:=H\phi_0\seq   
              \textbf{measure } \phi \; b_0b_1 
\end{align*}

That is, Alice applies one of the three Pauli operators or an identity
to her half of the entangled pair, depending on the values of her
classical bits, and sends her qubit to Bob. Bob receives the qubit,
applies a controlled-not followed by a Hadamard, and measures the two
qubits in his possession. We now show that the program $P$ implements
the specification $S$. First, we simplify the processes $Alice$ and
$Bob$:
\begin{align*}
  Alice \eqv & \phi_0:=(-i)^{a_0 \times a_1} \times Z^{a_0} (X^{a_1} \phi_0)\seq 
               q:=q+1\seq  qch! \phi_0 
            &\text{(math)}\\
  Bob \eqv  & qch? \phi_0\seq  \textbf{measure } H\tensor I (CNOT \phi) \; b_0b_1 
            &\hspace{-3mm}\text{(substitutions)}
\end{align*}

We now look at their parallel composition:
\begin{align*}
  P
\eqv 
  & \textbf{qchan } qch : qbit \cdot 
         Alice_{a_0,a_1,\phi_0} \; ||_{\phi}\;  Bob_{b_0,b_1,\phi_1}
  &\text{} \br
\eqv
  & \textbf{qchan } qch : qbit \cdot \\
   & ((\phi_0:= (-i)^{a_0 \times a_1} \times Z^{a_0} (X^{a_1} \phi_0)\seq 
       q:=q+1\seq  qch! \phi_0)_{a_0,a_1,\phi_0}\\
   & \; ||_{\phi}\;
     (qch? \phi_0; 
      \textbf{measure } H\tensor I (CNOT \phi) \; b_0b_1)_{b_0,b_1,\phi_1})
    &\\
 & \text{\qquad quantum channel}\br 
\eqv
   & \phi:=(-i)^{a_0 \times a_1} \times Z^{a_0}\tensor I (X^{a_1}\tensor I \phi)\seq \\
   & q'=q+1 \land 
     \textbf{var}_{Alice}' = \textbf{var}_{Alice} \backslash \phi_0 \land
     \textbf{var}_{Bob}' = \textbf{var}_{Bob},\phi_0 \land
     \sigma' = \sigma\seq \\
   & \textbf{measure } H\tensor I (CNOT\; \phi) \; b_0b_1\br
   &\text{\qquad sequential composition} \br
\eqv
   & (\textbf{measure } (-i)^{a_0 \times a_1} \times 
       H\tensor I (CNOT \;(Z^{a_0}\tensor I (X^{a_1}\tensor I \phi))) \; b_0b_1)
     \times \\
   & (q'=q+1) \times 
     (\textbf{var}_{Alice}' = a_0,a_1) \times
     (\textbf{var}_{Bob}' = b_0,b_1,\psi_0,\psi_1) \times
     (\sigma' = \sigma)
\end{align*}

Next, we note that the quantum state being measured is:
 \begin{align*}
    & (-i)^{a_0 \times a_1} \times 
      H\tensor I (CNOT\; (Z^{a_0}\tensor I (X^{a_1}\tensor I 
       (\ket{00}+\ket{11})/\sqrt{2})))
    &\text{}\br
 \eqv
    & (a_0=0)\times(a_1=0)\times
      H\tensor I (CNOT\; (\ket{00}+\ket{11})/\sqrt{2}) + \\
    & (a_0=0)\times(a_1=1)\times
      H\tensor I (CNOT\; (X\tensor I (\ket{00}+\ket{11})/\sqrt{2})) + \\
    & (a_0=1)\times(a_1=0)\times
      H\tensor I (CNOT\; (Z\tensor I (\ket{00}+\ket{11})/\sqrt{2})) + \\
    & (a_0=1)\times(a_1=1)\times
      (-i)\times 
      H\tensor I (CNOT\; (Z\tensor I (X\tensor I 
       (\ket{00}+\ket{11})/\sqrt{2})))\\
    &\qquad\text{apply X}\br
 \eqv
    & (a_0=0)\times(a_1=0)\times
      H\tensor I (CNOT\; (\ket{00}+\ket{11})/\sqrt{2}) + \\
    & (a_0=0)\times(a_1=1)\times
      H\tensor I (CNOT\; (\ket{10}+\ket{01})/\sqrt{2}) + \\
    & (a_0=1)\times(a_1=0)\times
      H\tensor I (CNOT\; (Z\tensor I (\ket{00}+\ket{11})/\sqrt{2})) + \\
    & (a_0=1)\times(a_1=1)\times
      (-i)\times 
      H\tensor I (CNOT\; (Z\tensor I  
       (\ket{10}+\ket{01})/\sqrt{2}))\\
    &\qquad\text{apply Z}\br
 \eqv
    & (a_0=0)\times(a_1=0)\times
      H\tensor I (CNOT\; (\ket{00}+\ket{11})/\sqrt{2}) + \\
    & (a_0=0)\times(a_1=1)\times
      H\tensor I (CNOT\; (\ket{10}+\ket{01})/\sqrt{2}) + \\
    & (a_0=1)\times(a_1=0)\times
      H\tensor I (CNOT\; (\ket{00}-\ket{11})/\sqrt{2}) + \\
    & (a_0=1)\times(a_1=1)\times
      (-i)\times 
      H\tensor I (CNOT\; (-\ket{10}+\ket{01})/\sqrt{2})\\
    &\qquad\text{apply CNOT}\br
 \eqv
    & (a_0=0)\times(a_1=0)\times
      H\tensor I (\ket{00}+\ket{10})/\sqrt{2} + \\
    & (a_0=0)\times(a_1=1)\times
      H\tensor I (\ket{11}+\ket{01})/\sqrt{2} + \\
    & (a_0=1)\times(a_1=0)\times
      H\tensor I (\ket{00}-\ket{10})/\sqrt{2} + \\
    & (a_0=1)\times(a_1=1)\times
      (-i)\times 
      H\tensor I (-\ket{11}+\ket{01})/\sqrt{2}\\
    &\qquad\text{apply H}\br
 \eqv
    & (a_0=0)\times(a_1=0)\times \ket{00} + \\
    & (a_0=0)\times(a_1=1)\times \ket{01} + \\
    & (a_0=1)\times(a_1=0)\times \ket{10}) + \\
    & (a_0=1)\times(a_1=1)\times (-i)\times\ket{11}\\
    \eqv
    & (-i)^{a_0 \times a_1} \times \ket{a_0a_1}
 \end{align*}

Putting it all together, we get:
\begin{align*}
  & (\phi_{01} = (\ket{00} + \ket{11})/\sqrt{2}) \land P 
  &\text{as above} \\
\eqv 
  & (\textbf{measure } (-i)^{a_0 \times a_1} \times \ket{a_0a_1} \; b_0b_1)
     \times (q'=q+1) \times\\
   & (\textbf{var}_{Alice}' = a_0,a_1) \times
     (\textbf{var}_{Bob}' = b_0,b_1,\psi_0,\psi_1) \times
     (\sigma' = \sigma) 
     &\text{measure} \\
  \eqv
  & (\phi'=\ket{b_0'b_1'}) \times (b_0'=a_0) \times (b_1'=a_1)
     \times (q'=q+1) \times \\
   & (\textbf{var}_{Alice}' = a_0,a_1) \times
     (\textbf{var}_{Bob}' = b_0,b_1,\psi_0,\psi_1) \times
     (\sigma' = \sigma)\\
   \leqv &S
\end{align*}

This example shows formalisation and analysis of a quantum
communication protocol which involves a quantum communication channel.


\section {Conclusion and Future Work}
\label{sec:conclusion}

We have presented a formal framework for specifying, implementing, and
analysing quantum communication protocols. The analysis is not
limited to reasoning about the data sent or received during the
execution of the protocol. We provide tools to formally prove
complexity of the communication protocols, such as the number of
classical and quantum bits sent during the execution. We have applied
our approach to two important quantum communication protocols: quantum
teleportation and quantum dense coding. The resulting formal proofs
are short: in fact, the proofs in Sections~\ref{sec:teleportation}
and~\ref{sec:densecoding} are only slightly longer than the informal
reasoning and calculations in~\cite{nielsen00}. The proofs are easy to
read, the use of Dirac-like notation makes the expressions of quantum
states look familiar, while providing a formal treatment that fits in
the overall framework. Finally, the formal proofs are checkable by a
computer (although we currently do not have suitable software
implemented), thus providing a measure of confidence in the analysis
of correctness and complexity of the protocols.

Current research focuses on formal reasoning about complexity of
distributed quantum algorithms (e.g.~\cite{yimsiriwattana04}). Future
work involves formalising quantum cryptographic protocols, such as
BB84~\cite{bennet84}, in our framework and providing formal analysis
of these protocols.


\bibliographystyle{entcs}
\bibliography{pqc}

\begin{thebibliography}{10}
\expandafter\ifx\csname url\endcsname\relax
  \def\url#1{\texttt{#1}}\fi
\expandafter\ifx\csname urlprefix\endcsname\relax\def\urlprefix{URL }\fi
\newcommand{\enquote}[1]{``#1''}

\bibitem{abramsky04}
Abramsky, S., \emph{High-level methods for quantum computation and
  information}, in: \emph{Proceedings of the 19th Annual IEEE Symposium on
  Logic in Computer Science} (2004), pp. 410--414.

\bibitem{abramsky04:csqp}
Abramsky, S. and B.~Coecke, \emph{A categorical semantics of quantum
  protocols}, in: \emph{Proceedings of the 19th Annual IEEE Symposium on Logic
  in Computer Science} (2004), pp. 415--425.

\bibitem{abramsky06}
Abramsky, S. and R.~Duncan, \emph{A categorical quantum logic}, Mathematical
  Structures in Computer Science \textbf{16} (2006), pp.~469--489.

\bibitem{adao}
Ad{\~{a}}o, P. and P.~Mateus, \emph{A process algebra for reasoning about
  quantum security}, in: \emph{Proceedings of the 3rd International Workshop on
  Quantum Programming Languages}, 2007, pp. 3--21.

\bibitem{altenkirch05:fqpl}
Altenkirch, T. and J.~Grattage, \emph{A functional quantum programming
  language}, in: \emph{Proceedings of the 20th Annual IEEE Symposium on Logic
  in Computer Science} (2005), pp. 249--258.

\bibitem{arrighi04}
Arrighi, P. and G.~Dowek, \emph{Operational semantics for formal tensorial
  calculus}, in: \emph{Proceedings of the 2nd International Workshop on Quantum
  Programming Languages}, 2004, pp. 21--38.

\bibitem{arrighi05}
Arrighi, P. and G.~Dowek, \emph{Linear-algebraic $\lambda$-calculus},
  arXiv:quant-ph/0501150 (2005).

\bibitem{bennet84}
Bennett, C.~H. and G.~Brassard, \emph{Quantum cryptography: Public-key
  distribution and coin tossing}, in: \emph{Proceedings of IEEE International
  Conference on Computers, Systems and Signal Processing}.

\bibitem{bennett93}
Bennett, C.~H., G.~Brassard, C.~Cr\'epeau, R.~Jozsa, A.~Peres and W.~K.
  Wootters, \emph{Teleporting an unknown quantum state via dual classical and
  {E}instein-{P}odolsky-{R}osen channels}, Phys. Rev. Lett. \textbf{70} (1993),
  pp.~1895--1899.

\bibitem{coecke04:le}
Coecke, B., \emph{The logic of entanglement}, arXiv:quant-ph/0402014 (2004).

\bibitem{danos07}
Danos, V., E.~Kashefi and P.~Panangaden, \emph{The measurement calculus},
  Journal of the ACM \textbf{54}.

\bibitem{dHondt06:wp}
D'Hondt, E. and P.~Panangaden, \emph{Quantum weakest preconditions},
  Mathematical Structures in Computer Science \textbf{16} (2006), pp.~429--451.

\bibitem{gay05}
Gay, S.~J. and R.~Nagarajan, \emph{Communicating quantum processes}, in:
  \emph{Proceedings of the 32nd ACM SIGACT-SIGPLAN Symposium on Principles of
  Programming Languages} (2005), pp. 145--157.

\bibitem{hehner:book}
Hehner, E.~C., \enquote{a Practical Theory of Programming,} Springer, New York,
  1993, first edition, current edn. (2008) Available free
  at~\url{www.cs.utoronto.ca/~hehner/aPToP}.

\bibitem{hehner04}
Hehner, E.~C., \emph{Probabilistic predicative programming}, in:
  \emph{Proceedings of the 7th International Conference on Mathematics of
  Program Construction},  Lecture Notes in Computer Science  \textbf{3125}
  (2004), pp. 169--185.

\bibitem{jorrand04}
Jorrand, P. and M.~Lalire, \emph{Toward a quantum process algebra}, in:
  \emph{Proceedings of the 1st ACM Conference on Computing Frontiers} (2004),
  pp. 111--119.

\bibitem{lalire04}
Lalire, M. and P.~Jorrand, \emph{A process algebraic approach to concurrent and
  distributed computation: operational semantics}, in: \emph{Proceedings of the
  2nd International Workshop on Quantum Programming Languages}, 2004, pp.
  109--126.

\bibitem{nielsen00}
Nielsen, M.~A. and I.~L. Chuang, \enquote{Quantum Computation and Quantum
  Information,} Cambridge University Press, 2000.

\bibitem{sanders00}
Sanders, J.~W. and P.~Zuliani, \emph{Quantum programming}, in:
  \emph{Mathematics of Program Construction},  Lecture Notes in Computer
  Science  \textbf{1837} (2000).

\bibitem{selinger04}
Selinger, P., \emph{Towards a quantum programming language}, Mathematical
  Structures in Computer Science \textbf{14} (2004), pp.~527--586.

\bibitem{tafliovich04}
Tafliovich, A., \enquote{Quantum Programming,} Master's thesis, University of
  Toronto (2004).

\bibitem{tafliovich06}
Tafliovich, A. and E.~C. Hehner, \emph{Quantum predicative programming}, in:
  \emph{Proceedings of the 8th International Conference on Mathematics of
  Program Construction},  Lecture Notes in Computer Science  \textbf{4014}
  (2006), pp. 433--454.

\bibitem{tafliovich07}
Tafliovich, A. and E.~C. Hehner, \emph{Programming telepathy: Implementing
  quantum non-locality games}, in: \emph{Proceedings of the 10th Brazilian
  Symposium on Formal Methods} (2007), pp. 70--86.

\bibitem{valiron04}
Valiron, B., \emph{Quantum typing}, in: \emph{Proceedings of the 2nd
  International Workshop on Quantum Programming Languages}, 2004, pp. 163--178.

\bibitem{yimsiriwattana04}
Yimsiriwattana, A. and S.~J.~L. Jr, \emph{Distributed quantum computing: A
  distributed {S}hor algorithm}, arXiv:quant-ph/0403146 (2004).

\bibitem{zuliani:thesis}
Zuliani, P., \enquote{Quantum Programming,} D{P}hil thesis, University of
  Oxford (2001).

\bibitem{zuliani04}
Zuliani, P., \emph{Non-deterministic quantum programming}, in:
  \emph{Proceedings of the 2nd International Workshop on Quantum Programming
  Languages}, 2004, pp. 179--195.

\end{thebibliography}

\appendix

\section {Quantum Computation}
\label{sec:qc}

In this section we introduce the basic concepts of quantum mechanics,
as they pertain to the quantum systems that we consider for quantum
computation. The discussion of the underlying physical processes,
spin-$\frac{1}{2}$-particles, etc.~is not our interest.  We are
concerned with the model for quantum computation only.  A reader not
familiar with quantum computing can consult~\cite{nielsen00} for a
comprehensive introduction to the field.

The \emph{Dirac notation}, invented by Paul Dirac, is often used in
quantum mechanics. In this notation a vector $v$ (a column vector by
convention) is written inside a \emph{ket}: $\ket{v}$.  The dual
vector of $\ket{v}$ is $\bra{v}$, written inside a \emph{bra}. The
inner products are \emph{bra-kets} $\braket{v}{w}$.  For
$n$-dimensional vectors $\ket{u}$ and $\ket{v}$ and $m$-dimensional
vector $\ket{w}$, the value of the inner product $\braket{u}{v}$ is a
scalar and the outer product operator $\ket{v}\bra{w}$ corresponds to
an $m$ by $n$ matrix.  The Dirac notation clearly distinguishes
vectors from operators and scalars, and makes it possible to write
operators directly as combinations of bras and kets.

In quantum mechanics, the vector spaces of interest are the Hilbert
spaces of dimension $2^n$ for some $n \in \mathbb{N}$.  A convenient
orthonormal basis is what is called a \emph{computational basis}, in
which we label $2^n$ basis vectors using binary strings of length $n$
as follows: if $s$ is an $n$-bit string which corresponds to the
number $x_s$, then $\ket{s}$ is a $2^n$-bit (column) vector with $1$
in position $x_s$ and $0$ everywhere else. The tensor product $\ket{i}
\tensor \ket{j}$ can be written simply as $\ket{ij}$.  An arbitrary
vector in a Hilbert space can be written as a weighted sum of the
computational basis vectors.

\begin{description}
\item[Postulate 1 (state space)] Associated to any isolated physical
  system is a Hilbert space, known as the \emph{state space} of the
  system. The system is completely described by its \emph{state
    vector}, which is a unit vector in the system's state space.
\end{description}

\begin{description}
\item [Postulate 2 (evolution)] The evolution of a closed quantum
  system is described by a \emph{unitary transformation}.
\end{description}

\begin{description}
\item [Postulate 3 (measurement)] Quantum measurements are described
  by a collection $\{M_m\}$ of \emph{measurement operators}, which act
  on the state space of the system being measured. The index $m$
  refers to the possible measurement outcomes. If the state of the
  system immediately prior to the measurement is described by a vector
  $\ket{\psi}$, then the probability of obtaining result $m$ is
  $\braket{\psi}{M_m^{\dagger} M_m |\psi}$, in which case the state of
  the system immediately after the measurement is described by the
  vector $\frac{M_m \ket{\psi}}{\sqrt{\braket{\psi}{M_m^{\dagger} M_m
        |\psi}}}$. The measurement operators satisfy the
  \emph{completeness equation} $\sum m \cdot M_m^{\dagger} M_m \eqv I$.
\end{description}

An important special class of measurements is \emph{projective
  measurements}, which are equivalent to general measurements provided
that we also have the ability to perform unitary transformations.

A projective measurement is described by an \emph{observable} $M$,
which is a Hermitian operator on the state space of the system being
measured. This observable has a spectral decomposition $M=\sum m \cdot
\lambda_m \times P_m$, where $P_m$ is the projector onto the
eigenspace of $M$ with eigenvalue $\lambda_m$, which corresponds to
the outcome of the measurement.  The probability of measuring $m$ is
$\braket{\psi}{P_m | \psi}$, in which case immediately after the
measurement the system is found in the state $\frac{P_m
  \ket{\psi}}{\sqrt{\braket{\psi}{P_m | \psi}}}$.

Given an orthonormal basis $\ket{v_m}$, $0 \leq m < 2^n$, measurement
with respect to this basis is the corresponding projective measurement
given by the observable $M = \sum m \cdot \lambda_m \times P_m$, where
the projectors are $P_m = \ket{v_m}\bra{v_m}$.

Measurement with respect to the computational basis is the simplest
and the most commonly used class of measurements. In terms of the
basis $\ket{m}$, $0 \leq m < 2^n$, the projectors are $P_m =
\ket{m}\bra{m}$ and $\braket{\psi}{P_m | \psi} = |\psi_m|^2$. The
state of the system immediately after measuring $m$ is $\ket{m}$.

For example, measuring a single qubit in the state $\alpha \times
\ket{0} + \beta \times \ket {1}$ results in the outcome $0$ with
probability $|\alpha|^2$ and outcome $1$ with probability $|\beta|^2$.
The state of the system immediately after the measurement is $\ket{0}$
or $\ket{1}$, respectively.

Suppose the result of the measurement is ignored and we continue the
computation. In this case the system is said to be in a \emph{mixed
  state}. A mixed state is not the actual physical state of the
system. Rather it describes our knowledge of the state the system is
in.  In the above example, the mixed state is expressed by the
equation $\ket{\psi} = |\alpha|^2 \times \{\ket{0}\} + |\beta|^2
\times \{\ket{1}\}$. The equation is meant to say that $\ket{\psi}$ is
$\ket{0}$ with probability $|\alpha|^2$ and it is $\ket{1}$ with
probability $|\beta|^2$. An application of operation $U$ to the mixed
state results in another mixed state, $U(|\alpha|^2 \times \{\ket{0}\}
+ |\beta|^2 \times \{\ket{1}\}) = |\alpha|^2 \times \{U\ket{0}\} +
|\beta|^2 \times \{U\ket{1}\}$.

\begin{description}
\item [Postulate 4 (composite systems)] The state space of a composite
  physical system is the tensor product of the state spaces of the
  component systems. If we have systems numbered $0$ up to and
  excluding $n$, and each system $i$, $0 \leq i < n$, is prepared in
  the state $\ket{\psi_i}$, then the joint state of the composite
  system is $\ket{\psi_0} \tensor \ket{\psi_1} \tensor \ldots \tensor
  \ket{\psi_{n-1}}$.
\end{description}

While we can always describe a composite system given descriptions of
the component systems, the reverse is not true. Indeed, given a state
vector that describes a composite system, it may not be possible to
factor it to obtain the state vectors of the component systems. A
well-known example is the state $\ket{\psi} = \ket{00}/\sqrt{2} +
\ket{11}/\sqrt{2}$. Such a state is called an \emph{entangled} state.

Just as it may not be possible to represent the state of a multi-qubit
system as tensor product of its component systems, it may not be
possible to represent an operation on a composite system as a tensor
product of single-qubit operations on the component systems. Consider,
for example, ``controlled-NOT'' (CNOT) operation on two qubits defined
by
\begin{align*}
  &CNOT (\ket{0}\tensor\ket{x}) = \ket{0}\tensor\ket{x} \\
  &CNOT (\ket{1}\tensor\ket{x}) = \ket{1}\tensor\ket{1-x}
\end{align*}
where $x \in {0,1}$. It can be shown that there are no two
single-qubit operations $U_0$ and $U_1$, such that $CNOT = U_0 \tensor
U_1$.


\end{document}